\documentclass[12pt]{article}
\usepackage{a4wide}
\usepackage{amssymb}
\usepackage{graphicx}
\begin{document}
{\renewcommand{\thefootnote}{\fnsymbol{footnote}}
\begin{center}
{\LARGE  MONDified Gravity}\\
\vspace{1.5em}
Martin Bojowald\footnote{e-mail address: {\tt bojowald@psu.edu}}
and Erick I.\ Duque\footnote{e-mail address: {\tt eqd5272@psu.edu}}
\\
\vspace{0.5em}
Institute for Gravitation and the Cosmos,\\
The Pennsylvania State
University,\\
104 Davey Lab, University Park, PA 16802, USA\\
\vspace{1.5em}
\end{center}
}

\setcounter{footnote}{0}

\begin{abstract}
  A new class of modified gravity theories, made possible by subtle features
  of the canonical formulation of general covariance, naturally allows
  MOND-like behavior (MOdified Newtonian Dynamics) in effective space-time
  solutions without introducing new fields. A detailed analysis reveals a
  relationship with various quantum-gravity features, in particular in
  canonical approaches, and shows several properties of potential
  observational relevance. A fundamental origin of MOND and a corresponding
  solution to the dark-matter problem are therefore possible and testable.
\end{abstract}

\section{Introduction}

Applications of general relativity to cosmology at low curvature and,
increasingly, to black holes in strong-field regimes have led to several
unexplained phenomena, highlighting the need to find alternative gravitational
theories for detailed comparisons with observations. The requirement that such
theories be generally covariant is often taken as implying that they
must be related to general relativity by introducing additional interactions
from higher-curvature terms or from new fields of scalar, vector or tensor
nature. However, many such theories are ruled out by the observational insight
that the speed of gravitational waves is very close to the speed of light
\cite{MultiMess1,MultiMess2,MultiMess3,MultiMess4}. Moreover, higher-curvature
actions often have instabilities caused by higher time derivatives
\cite{OstrogradskiProblem}.

Recent results in canonical gravity \cite{SphSymmEff,HigherCov} have
culminated in the conclusion that the usual road that leads to
higher-curvature or scalar-vector-tensor actions is not the only one to
alternative gravity theories, thanks to a subtle feature of general
covariance: Its mathematical formulation, expressed canonically by conditions
on fields on a foliation of space-time into spacelike hypersurfaces, does not
take the same form as in the common picture of coordinate changes in a
4-dimensional space-time manifold. Conditions that ensure general covariance
of the foliation have been constructed early on in canonical formulations of
general relativity \cite{DiracHamGR}, requiring the imposition of constraints
which generate hypersurface deformations as a gauge symmetry. It is well-known
that algebraic properties of the constraints are rather complicated because
they have Poisson brackets or commutators with structure functions that depend
on the fundamental fields, in particular on the spatial metric on a
hypersurface. (In an independent line of mathematical research, the
relationship between this feature and an $L_{\infty}$-structure that modifies
the usual Jacobi identity has been analyzed in \cite{ConsRinehart}.)

Classically, the structure function equals the inverse of a fundamental
spatial metric field used to define the theory, and all standard theories of
modified gravity in metric form maintain this relationship. The opening to new
classes of modified gravity consists in the observation that the structure
function may well have a different relationship with the fundamental
fields. Provided that its gauge transformations take the form required for
coordinate transformations of an inverse spatial metric, the structure
function then defines an {\em emergent} space-time line element distinct from
the fundamental fields. With hindsight, the results of
\cite{SphSymmEff} show that there are indeed new non-trivial
candidates for this new class of {\em emergent modified gravity}, at least in
spherically symmetric models. Here, we provide a complete fundamental
formulation as well as new applications.

In particular, we use our new formulation of the underlying canonical theory
to show that several of its features can imply a natural relativistic
realization of Modified Newtonian Dynamics (MOND, \cite{MOND1,MOND2}) without
introducing extra fields. Constructing such theories has proven difficult in
manifestly covariant form in a standard space-time description, unless new
non-geometrical fields are introduced \cite{TEVES,TEVES2,TEVES3}. The new
theories discussed here may therefore help to analyze or explain long-standing
cosmological puzzles such as dark matter. Their fundamental nature is
highlighted by several examples provided here that show how terms required for
emergent modified gravity may be implied by ingredients of canonical quantum
gravity. We discuss properties of orbital motion and light deflection in
emergent modified gravity in order to show how its implication can be
confronted by observations.

\section{Emergent modified gravity}

Infinitesimal deformations of spacelike hypersurfaces in space-time are known
to have an unusual and challenging feature: While infinitesimal deformations
$D(\vec{M})$ along a spatial vector field $\vec{M}$ tangential to a
spacelike hypersurface have a simple commutator
\begin{equation}\label{DD}
  [ D ( \vec{M}_1) , D ( \vec{M}_2 ) ] = D
  (\mathcal{L}_{\vec{M}_1} \vec{M}_2)
\end{equation}
given by a directional or Lie derivative $\mathcal{L}_{\vec{M}_1} \vec{M}_2$,
the commutator of two deformations $H(N)$ normal to the hypersurface by a
displacement function $N$,
\begin{equation}\label{HH}
  [ H ( N_1 ) , H ( N_2 ) ] = D ( N_1 \vec{\nabla} N_2 - N_2\vec{\nabla}N_1 ))\,,
\end{equation}
depends, through the gradient $\nabla^aN=q^{ab}\partial_bN$, on the metric
$q_{ab}$ induced on the hypersurface, with inverse $q^{ab}$. Since the
hypersurfaces may be curved with varying spatial geometries, the commutator
depends on the hypersurface to which it is applied. The generators
$D(\vec{M})$ and $H(N)$ of gauge transformations are constrained to vanish on
physical solutions of the theory. The presence of structure functions then
complicates quantizations of these equations because it requires specific
factor orderings of non-commuting metric and constraint operators. In the
context of classical modified gravity, however, the presence of a structure
function is powerful because it gives us a direct link from algebraic
properties of gauge transformations to geometrical structures of space or
space-time.

We analyze the specific form of the constraints for spherically symmetric line
elements of the general form
\begin{equation} \label{ds}
  {\rm d}s^2= -N^2{\rm d}t^2+ \frac{(E^{\varphi})^2}{E^x} ({\rm d}x+N^x {\rm
    d}t)^2+ E^x({\rm d}\vartheta^2+ \sin^2\vartheta{\rm d}\varphi^2)
\end{equation}
in which all functions depend only on $x$ and $t$. The functions $N$ and $N^x$
parameterize foliations of space-time into spacelike hypersurfaces, on which
the induced metric is described by two functions, $E^{\varphi}$ and
$E^x$. These two functions, in a canonical formulation, are the momenta of
components of extrinsic curvature, $K_{\varphi}$ and $K_x$ \cite{SphSymm}. In
canonical form, candidates for theories of gravity are given by phase-space
functions $D$ and $H$ with Poisson brackets of the form (\ref{DD}) and
(\ref{HH}), replacing the action functional. In general, the classical
$q^{ab}$ in $\nabla^aN$, or $q^{xx}$ in spherical symmetry, is then some
phase-space function $\tilde{q}^{xx}(E^x,E^{\varphi},K_x,K_{\varphi})$ that is
not required to have a simple relationship with the fundamental canonical
fields. A candidate theory defined like this is generally covariant if the
structure function $\tilde{q}^{xx}$ is subject to gauge transformations
consistent with coordinate changes of an inverse spatial metric. It can then
be used to define a compatible emergent space-time line element, supplying a
geometrical interpretation to solutions of the theory.

The covariance conditions can be evaluated in a canonical version of effective
field theory, in which one starts with a general expression for the integrand
of $H(N)=\int{\rm d}x N H$ as a local function of the canonical fields up to
some order in spatial derivatives and a certain polynomial order in $K_x$,
which has a non-zero density weight. The coefficients of spatial derivatives
of $E^x$ and $E^{\varphi}$ and of the $K_x$-terms are initially free functions
of $E^x$ and $K_{\varphi}$ (which both have zero density weight), but
their form is strongly restricted by two conditions, that (i) the Poisson
bracket $\{H(N_1),H(N_2)\}$ be of the form (\ref{HH}) with (ii) a structure
function $\tilde{q}^{xx}$ that has gauge transformations consistent with
coordinate transformations of an inverse spatial metric.

As shown in \cite{HigherCov}, completing earlier versions in
\cite{SphSymmMatter,SphSymmMatter2}, the most general such expression for a
generator $H$ of normal deformations takes the form\footnote{A similar version
  was presented in \cite{SphSymmMinCoup} while the journal version of this
  paper was being published.}
\begin{eqnarray}
    H
    &=&  - \lambda_0 \frac{\sqrt{E^x}}{2} E^\varphi \Bigg[
    c_{f 0}
    + \frac{\alpha_0}{E^x}
    + 2 \frac{\sin^2 \left(\bar{\lambda}
        K_\varphi\right)}{\bar{\lambda}^2}\frac{\partial c_{f}}{\partial E^x} 
    + 4 \frac{\sin \left(2 \bar{\lambda} K_\varphi\right)}{2 \bar{\lambda}}
        \frac{\partial q}{\partial E^x} 
    \nonumber\\
    &&\qquad\qquad\quad + \frac{\alpha_2}{\bar{\lambda}^2 E^x} c_f- \frac{(E^\varphi)^2}{\lambda_0^2 \bar{\lambda}^2 E^x} \left(
       \frac{\alpha_2}{E^x} \tilde{q}^{x x} 
    + 2 \frac{K_x}{E^\varphi} \frac{\partial \tilde{q}^{x x}}{\partial
       K_\varphi}\right)\nonumber\\
&&  \qquad\qquad\quad
    - \left( \frac{(E^x)' \left((E^\varphi)^{- 2}\right)'}{2}   + \frac{(E^x)''}{(E^\varphi)^2} \right) \cos^2 \left( \bar{\lambda} K_\varphi \right)
    \Bigg]
    \label{H}
\end{eqnarray}
when it is expanded up to second order in spatial derivatives (using units
such that the speed of light and Newton's constant equal one), where
\begin{equation}
    \tilde{q}^{x x}
    =
    \left(
    \left( c_{f}
    + \left(\frac{\bar{\lambda} (E^x)'}{2 E^\varphi} \right)^2 \right) \cos^2 \left(\bar{\lambda} K_\varphi\right)
    - 2 q \bar{\lambda}^2 \frac{\sin \left(2 \bar{\lambda} K_\varphi\right)}{2 \bar{\lambda}}\right)
    \lambda_0^2 \frac{E^x}{(E^\varphi)^2}
    \label{q}
\end{equation}
is the structure function replacing $q^{xx}=E^x/(E^{\varphi})^2$ in the
classical bracket (\ref{HH}).  Here, $\bar{\lambda}$ is a free constant, and
$\lambda_0$, $c_f$, $c_{f0}$, $\alpha_0$, $\alpha_2$ and $q$ are free
functions of $E^x$ not determined by the covariance condition alone. The free
parameters take simple values zero or one in classical general relativity,
which demonstrates the existence of a number of new modified gravity
theories. (More precisely, we obtain the classical limit for
$\lambda_0 , c_f , \alpha_1, \alpha_2 \to 1$, $\bar{\lambda} \to 0$, and
$c_{f0} \to - \Lambda$ if there is a cosmological constant.) Among the
  free functions, only $c_f$, $\lambda_0$ and $q$ (for $\bar{\lambda}\not=0$)
  appear in the emergent spatial metric and are therefore characteristic of
  emergent modified gravity, while $c_{f0}$, $\alpha_0$ and $\alpha_2$
  parameterize freedom present in the classical spherically symmetric theory,
  akin to the free potential of dilaton gravity. The phase space
is not enlarged, and therefore there are no additional fields such as those
implied by higher time derivatives in a canonical formulation, or additional
scalar, vector or tensor fields added to general relativity by hand.

The rather involved relationship (\ref{q}) is unambiguously determined by the
Poisson bracket of two modified Hamiltonian constraints of the form
(\ref{HH}). The fact that $\tilde{q}^{xx}$ is not a simple combination of the
basic phase-space functions demonstrates that these theories are not easily
related to action principles formulated for a fundamental space-time metric,
as done in standard modified gravity. Nevertheless, dynamical and gauge
properties, and therefore physical predictions, are uniquely determined by
Hamilton's equations generated by the constraints. 

Given the canonical structure and the covariance condition, it is possible to
evaluate the space-time dynamics implied by $H$ and the corresponding emergent
line element with inverse spatial metric (\ref{q}). The freedom of choosing a
suitable space-time slicing allows us to impose additional conditions in order
to simplify the calculations or to focus attention on specific space-times
such as static ones close to the standard Schwarzschild solution. For the
latter, we choose $E^x=x^2$ and $N^x=0$. (According to (\ref{ds}), the radial
coordinate $x$ then determines the area of symmetric spheres by the usual
equation $A=4\pi x^2$.)

The second choice, $N^x=0$, implies that the equations of motion are given by
Hamilton's equations of $NH$ (rather than the full $H(N)+D(N^x)$), where
$N$ is the same space-time function that appears in (\ref{ds}).  The first
choice, $E^x=x^2$, implies a consistency condition because $E^x$ is then
time-independent and the corresponding Hamilton equation must be zero. Given
the general Hamiltonian, this condition reads
$\dot{E}^x=-N\partial H/\partial K_x=0$, or
$\tan \left(2 \bar{\lambda} K_\varphi\right) = - 2\bar{\lambda} q/( c_{f} +
\bar{\lambda}^2 x^2/(E^\varphi)^2)$.  A simple solution of $K_{\varphi}=0$ is
obtained if we choose a modified dynamics with the classical value $q=0$.  The
other curvature $K_x=0$, then also vanishes as a consequence of $D=0$. The
constraint $H=0$ implies
\begin{equation}
    0 =
    (\ln (E^\varphi)^2)'
    + \frac{(E^\varphi)^2}{x} \left(
    c_{f 0}
    + \frac{\alpha_0}{x^2} \right)
    - \frac{\alpha_2 + 2}{x}
    \,,
    \label{Ephi}
\end{equation}
which for specific choices of $c_{f0}$ and $\alpha_2$ can be solved for
$E^{\varphi}$. The remaining Hamilton equations are consistent with a static
solution, $\dot{E}^{\varphi}=0$, and a final condition,
\begin{equation}
    \left(\ln ( N E^\varphi) \right)'
    =
    \frac{\alpha_2}{x}
    - (\ln \lambda_0)'
    \,,
    \label{N}
\end{equation}
is implied by $\dot{K}_{\varphi}=0$ because $K_{\varphi}=0$. The condition
$\dot{K}_x=0$ does not result in an independent equation.

Equation~(\ref{Ephi}) may be solved by making an ansatz
$E^\varphi(x) = x/\sqrt{1 - f_\varphi(x)}$ with a function $f_{\varphi}$
subject to the equation
\begin{equation}
    0 =
    xf_{\varphi}'
    + \alpha_2 f_{\varphi}
    + \alpha_0 - \alpha_2
    + c_{f0} x^2
    \,.
    \label{fphieq}
\end{equation}
If the free functions are such that the classical limit with a vanishing
cosmological constant is obtained for large $x$, the solution to this equation
is asymptotically of the form $f_{\varphi}(x)\sim 2M/x$ with a constant $M$
that is interpreted as the central mass, such as a black hole, that gives rise
to our curved space-time.
The solution of (\ref{N}) for $N$ then takes the form
\begin{equation}
    N(x) = \frac{\sqrt{1 - f_\varphi(x)}}{\mu(x) \lambda_0(x)}
    \ ,
    \label{Nsol}
\end{equation}
provided
\begin{equation}
    (\ln \mu)'
    =
    \frac{1- \alpha_2(x)}{x}
    \,.
    \label{mueq}
  \end{equation}

These solutions simplify (\ref{q}) and imply 
an emergent space-time line element
\begin{equation}
    {\rm d} s^2 =
    - \frac{\left(1-f_\varphi(x)\right)}{\mu(x)^2 \lambda_0(x)^2} {\rm d} t^2
    + \frac{{\rm d} x^2}{\lambda_0(x)^2
    \left( c_{f}(x)
    + \bar{\lambda}^2 \left(1-f_\varphi(x)\right) \right)
    \left(1-f_\varphi(x)\right)}
    + x^2 {\rm d} \Omega^2
    \label{dssol}
\end{equation}
which is clearly different from the general form of the Schwarzschild line
element, implying new effects.  A characteristic feature of (\ref{dssol}) is
that $g_{xx}\not=-g_{tt}^{-1}$ in general. Several new phenomena can be
expected from this inequality, for instance in the behavior of horizons, but
most of them depend on details of the specific modification functions $c_f$
and $\lambda_0$ and the corresponding solutions $f_{\varphi}$ and $\mu$. We
will focus instead on a more generic phenomenon that may show implications
even in non-relativistic regimes. To this end, we will first derive an
effective gravitational potential for geodesic motion of massive objects in a
space-time with line element (\ref{dssol}).

\section{Gravitational potential}

A standard procedure derives geodesic motion using spherical and time
translation symmetry of our space-time, together with the mass condition
$||{\bf p}||^2= g_{\mu \nu} p^\mu p^\nu = - m^2$ for the 4-momentum
${\bf p}=m{\rm d}{\bf x}/{\rm d}\tau$ of an object of mass $m$, using proper
time $\tau$. Killing vectors for the symmetries provide the conserved
energy $E=(1-f_{\varphi})\mu^{-2}\lambda_0^{-2} {\rm d}t/{\rm d}\tau$
and angular momentum $L=x^2{\rm d}\varphi/{\rm d}\tau$, and the mass condition
takes the form
\begin{equation}
    0 =
    \frac{1}{2} m \left(\frac{{\rm d}x}{{\rm d}\tau}\right)^2
    + \frac{\tilde{q}^{x x}}{2} \left( m
    + \frac{L^2}{mx^2}
    - \frac{E^2}{mN^2} \right)
\end{equation}
of a Newtonian-style energy balance, where $\tilde{q}^{xx}$ is the inverse
spatial metric and $-N^2$ the time component in (\ref{dssol}). Geodesic motion
of massive objects can therefore be expressed by the effective potential
\begin{equation}\label{Veffm}
  V(x)
  =
    \frac{\lambda_0^2}{2} \left( c_f
    + \bar{\lambda}^2 \left(1 - f_\varphi\right) \right)
    \left( 
    \left(1 - f_\varphi\right) \left( m
    + \frac{L^2}{mx^2} \right)
    - \frac{\mu^2 \lambda_0^2 E^2}{m} \right) \,.
\end{equation}

In the classical case, in which the $x$-dependent functions $\mu$, $c_f$ and
$\lambda_0$ equal one while $\bar{\lambda}=0$ and the solution to
(\ref{fphieq}) is $f_{\varphi}(x)=2M/x+\Lambda x^2$, (\ref{Veffm}) contains
Newton's gravitational potential, the centrifugal potential, and a
long-distance correction from $\Lambda$. Depending on the modification
functions, the potential may deviate from the two classical power laws of
$x^{-1}$ and $x^2$. In a region with small and constant deviations from the
classical values for the remaining parameters, $\alpha_2\approx 1$,
$\alpha_0\approx 1$, $c_{f 0} \approx - \Lambda$, and some $\lambda_0$, the
solution of (\ref{fphieq}) stays close to its classical form,
\begin{equation} \label{falpha}
    f_\varphi(x) \approx
    \frac{2 M}{x^{\alpha_2}} + \frac{\alpha_2-\alpha_0}{\alpha_2}x^{\alpha_2}
    + \frac{\Lambda}{3} \frac{1}{\alpha_2+2} x^{2+\alpha_2}
    \,,
\end{equation}
and $\mu$ depends only weakly on $x$ as a consequence of (\ref{mueq}). The
Newtonian power law may therefore change, but not significantly enough for
MOND-like effects that would require a logarithmic contribution to
$f_{\varphi}$ in order to result in a $1/x$ force law on intermediate scales.

Logarithmic MOND effects can, however, appear in the gravitational potential
via more dramatic modifications in at least two ways.  The first is by
choosing $\alpha_0=\alpha_2=1$ and
$c_{f 0} = - \Lambda + (c_1/E^x) \ln \left(e^2 E^x/c_2\right)$, with constants
$c_1$ and $c_2$ with vanishing $c_1$ in the classical limit, such that
(\ref{fphieq}) is solved by
$f_\varphi=2 M/x + \Lambda x^2/3 - c_1 \ln \left(x^2/c_2\right)$ despite
having $c_{f0} \approx - \Lambda$ at large scales.  The second, simpler way
that we will mainly focus on is by simply adding a logarithmic contribution to
the function $c_f$, which directly appears in the gravitational potential.
Our new theories of emergent modified gravity allow logarithmic terms in these
functions while preserving general covariance. More importantly, as we will
now show, there are several reasons for quantum-gravity effects to imply
logarithmic contributions to $c_f$, which appears directly in the
gravitational Hamiltonian and the structure function. Such terms would be
relevant on intermediate scales, between the Newtonian one dominated by $2M/x$
and the cosmological one dominated by $\Lambda x^2/3$. This is the extra-solar
or galactic range of MOND.

A well-known source of logarithmic terms in quantum field theories is given by
renormalization. In the present case of a canonical theory, this process
requires Hamiltonian renormalization in a background-independent manner, which
is still being developed; see for instance \cite{HamRen} or the moment
derivation of the Coleman--Weinberg potential in \cite{CW}, to which
renormalization can be applied as in \cite{ColemanWeinberg}. Treating
gravitational models in this way remains a challenge, but there are
independent results from \cite{SphSymmMoments} that explicitly show a
non-classical logarithmic term in the Hamiltonian constraint of a
quasiclassical spherically symmetric model, resulting from a solution for
quantum fluctuations of the metric components.  Importantly, these canonically
derived logarithms need not be Lorentz invariants, as in standard
renormalization theory. They cannot be implemented in standard
higher-curvature actions but may find a new home in emergent modified gravity.

We are also able to present an independent calculation that leads to
logarithmic terms in $c_f$, based on our new methods for the gravitational
constraints and the covariance condition \cite{HigherCov}. As suggested in
\cite{LoopSchwarz}, canonical quantization of a constrained system simplifies
if one can eliminate the structure function, given by $\tilde{q}^{xx}$ in
spherically symmetric models. This may be possible by considering suitable
linear combinations of the constraints with phase-space dependent
coefficients. The construction given in \cite{LoopSchwarz} can be generalized
significantly, while also making sure that the resulting theories remain local
and generally covariant. An example of such a solution obtained from a
covariant linear combination of the constraints without structure functions,
assuming the classical values $\alpha_0=1$ and $c_{f0}=-\Lambda$, is given by
\begin{equation}\label{cf}
  c_f(E^x) =1+\frac{\bar{\lambda}^2}{2} \left(\Lambda E^x- \ln(E^x/c_0)\right)
\end{equation}
where $c_0$ is a positive integration constant unrelated to the other
modification functions.

Since this modification is more significant on intermediate scales than the
constant $\alpha$-dependence in the function (\ref{falpha}), we can
approximate $f_{\varphi}(x)$ by its classical form. Up to a slowly-changing
factor of $\mu^{-1}\lambda_0^{-1}$, the lapse function (\ref{Nsol}) then also
has its classical form.  Nevertheless, the geometry is non-classical because
$c_f$ appears in the emergent line element (\ref{dssol}), now given by
\begin{eqnarray}
    {\rm d} s^2 &=&
    - \left( 1 - \frac{2 M}{x} - \frac{\Lambda}{3} x^2 \right) \frac{{\rm d}
                    t^2}{\mu^2\lambda_0^2}\nonumber\\ 
    &&+ \left( 1
    + \bar{\lambda}^2 \left(1
    - \frac{2 M}{x}
    - \ln \left(\frac{x}{\sqrt{c_0}}\right)
    + \frac{\Lambda}{6} x^2 \right) \right)^{-1} \frac{\lambda_0^{-2} {\rm d}
  x^2}{1 - 2 M/x - \Lambda x^2/3} \nonumber\\
    &&+ x^2 ({\rm d}\vartheta^2 + \sin^2\vartheta{\rm d}\varphi^2)
    \,.
    \label{eq:Spacetime - Schwarzschild gauge - Static - Partial Abelianization}
\end{eqnarray}
Independently of the origin of (\ref{cf}), our results show that there is a
modified gravity theory in canonical form which has this emergent line element
as a solution.

At distances large enough to neglect the Newtonian potential, but not so
large that the cosmological term is relevant, we have an effective
potential dominated by the logarithmic term. For non-relativistic objects with
$E\approx m$, we have
\begin{equation}
V \approx \frac{m \lambda_0^2}{2} \left( 1 + \bar{\lambda}^2 \left( 1 - \ln
    \left(\frac{x}{\sqrt{c_0}}\right) \right) \right)
\left(\frac{L^2}{m^2x^2}+ 1-\mu^2\lambda_0^2\right)\,.
\end{equation}
The combination of a negative logarithmic and a quadratic potential implies
new stable circular orbits at a radius $x_0$, which we may use for an estimate
of the relevant distance and velocity scales.  At a local minimum of the
potential, the angular momentum is given by
\begin{equation}
    \frac{L^2}{m^2} =
    \frac{\bar{\lambda}^2}{2} \left( \mu^2 \lambda_0^2 - 1 \right) \left( 1 - \frac{\bar{\lambda}^2}{2} \left( \ln \left(\frac{x^2}{c_0 e^2}\right) - 3 \right) \right)^{-1} x^2
    \,.
    \label{eq:Angular momentum - c_f MOND}
\end{equation}
In this condition, a stable circular orbit is then seen to exist only if
$\mu^2 \lambda_0^2 > 1$, such that, to leading order in the $\bar{\lambda}$
expansion, it has the radius
$x_0 = L(\bar{\lambda} m)^{-1} \sqrt{2/\mu^2 \lambda_0^2 - 1}$ and velocity
$v_0^2 = \frac{1}{2}\bar{\lambda}^2 (\mu^2 \lambda_0^2 - 1)$.
The Tully-Fisher relation \cite{MONDTest} $v_0^4 \approx a_0 M$, where
$a_0\approx \sqrt{\Lambda/3}/ (2 \pi)$ is Milgrom's universal acceleration
\cite{MONDRotation}, requires a mass-dependent
$\mu^2 \lambda_0^2 - 1 = 2 \sqrt{a_0 M} / \bar{\lambda}^2$.  In our
derivation, $\mu$ is part of the lapse function which usually depends on the
mass in static solutions, and $\lambda_0$ is expected to depend on $M$ if its
deviation from the classical value is implied by renormalization effects in
quantum gravity. Notice also that $\lambda_0$ multiplies the gravitational
Hamiltonian, just like Newton's constant does if gravity is coupled to matter,
and may therefore be renormalized.  The expression (\ref{eq:Angular momentum -
  c_f MOND}) implies that further MOND effects may be seen at even larger
scales.  However, because $\bar{\lambda}$ is expected to be small, such scales
may be cosmological and $\Lambda$ effects would have to be taken into account
too.

Since our theories of emergent modified gravity are fully relativistic and
generally covariant, it is straightforward to derive additional physical
effects such as modifications of the bending of light, which may be used to
subject the modification functions to further observational bounds.
For instance, the deflection angle of light moving around a central mass now
takes the form
\begin{equation}
    \Delta\phi\approx
    \frac{\pi}{\lambda_0} \left( 1
    - \frac{\sqrt{a_0 M}}{2}
    + \frac{\bar{\lambda}^2}{2} \ln \left(\frac{2 b}{\sqrt{c_0} e} \right)
  \right)
  \label{eq:Deflection angle - c_f MOND}
\end{equation}
to leading order in $a_0$ and $\bar{\lambda}$, with the impact parameter $b$
of the light ray.  Unlike the usually expected deflection angle in the MOND
literature which is independent of the impact parameter, there is a
logarithmic dependence in this relativistic extension and this could lead to a
negative deflection angle repelling the light.  In a semiclassical regime,
however, we expect $\sqrt{a_0 M} < \bar{\lambda}^2$, so the overall correction
to the deflection angle is positive for large enough impact parameters.

We briefly discuss the alternative MOND-like modification from
$c_{f 0}=-\Lambda+(c_1/E^x) \ln(e^2E^x/c_2)$. From a fundamental perspective,
this option is less preferred because it requires two undetermined constants
rather than only one. Moreover, because $c_{f0}$ does not appear in the
emergent spatial metric, it can be obtained also by choosing a suitable
dilaton potential. Phenomenologically, it also has more complicated and
ambiguous results.  Following a similar procedure as above it results in the
angular momentum expression
\begin{equation}
   \frac{L^2}{m^2 x^2}
    = \frac{ c_1\left(1 + \lambda^2
    + \lambda^2 \left(1 - \mu^2 \lambda_0^2\right)
    + 2 c_1 \lambda^2 \ln (x^2/c_2)\right)}
    {c_1^-
    + \lambda^2 \left( 2 c_1^--1 \right)
    + c_1 \left( 1 + 2 \lambda^2 c_1^- \right) \ln (x^2/c_2)
    + c_1^2 \lambda^2 \ln (x^2/c_2)^2}\label{eq:Angular momentum - fphi MOND}
\end{equation}
substituting (\ref{eq:Angular momentum - c_f MOND}) and using the shortcut
$c_1^-=1-c_1$.  The Tully-Fisher relation can be obtained in several ways.
For example, the choice $c_1 = \sqrt{a_0 M}$ (which needs renormalization for
$c_1$) reproduces the Tully-Fisher relation to zeroth order in
$\bar{\lambda}$.  Alternatively, the constant $\mu$ can be used in several
different ways and the expansions are ambiguous; one such realization is
$1 - \mu^2 \lambda_0^2 = \sqrt{a_0 M}/(c_1 \bar{\lambda}^2)$, which reproduces
the Tully-Fisher relation to zeroth order in both $c_1$ and $\bar{\lambda}$.
The deflection angle of light is instead, to leading order in $c_1$ and
$\bar{\lambda}$, given by
\begin{equation}
    \Delta \phi
    \approx
    \frac{\pi}{\lambda_0} \left(
    1
    - \bar{\lambda}^2 \left( c_1
    + \frac{1}{2} \right)
    - \frac{c_1}{2} \left(1+\frac{\bar{\lambda}^2}{2}\right) \ln \left( \frac{4 b^2}{c_2 e^2 \mu^2 \lambda_0^2} \right) \right)
    \ .
    \label{eq:Total change in angular coordinate - null ray - Static - MOND Alternative}
\end{equation}
If we take $c_1 = \sqrt{a_0 M}$ then this result differs from
(\ref{eq:Deflection angle - c_f MOND}) by a factor of approximately two when
taking the limit $\bar{\lambda}\to 0$ with fixed $a_0$.  Furthermore,
depending on the value of $c_2$, at large enough impact parameter the light is
only repelled.

These discrepancies can be used to distinguish between the two alternatives
using the data on deflection angles.  It is also possible to couple various
matter fields to gravity in our canonical formulation, and thereby investigate
the stability of compact objects.  Unlike previous proposals such as
\cite{TEVES,TEVES2,TEVES3}, our theory of emergent modified gravity makes it
possible to formulate MOND-like effects in a generally covariant form without
introducing additional fields or higher time derivatives. Stability is
therefore easier to ensure, and since there is a single emergent line element
that determines the propagation of all massless objects, gravitational waves
and light travel at the same speed.  The space-time properties of the horizon
and the interior region are well-behaved, and it is possible to obtain a
global structure with singularity resolution for certain values of the
constant $c_0$.  We therefore have a promising class of new observationally
viable alternatives to general relativity.

\section*{Acknowledgements}

This work was supported in part by NSF grant PHY-2206591.


\begin{thebibliography}{10}

\bibitem{MultiMess1}
LIGO Scientific, Virgo, Fermi GBM, INTEGRAL, IceCube, IPN, Insight-Hxmt,
  ANTARES, Swift, Dark Energy~Camera GW-EM, DES, DLT40, GRAWITA, Fermi-LAT,
  ATCA, ASKAP, OzGrav, DWF (Deeper Wider~Faster Program), AST3, CAASTRO,
  VINROUGE, MASTER, J-GEM, GROWTH, JAGWAR, CaltechNRAO, TTU-NRAO, NuSTAR,
  Pan-STARRS, KU, Nordic~Optical Telescope, ePESSTO, GROND, Texas~Tech
  University, TOROS, BOOTES, MWA, CALET, IKI-GW Follow-up, H.E.S.S., LOFAR,
  LWA, HAWC, Pierre Auger, ALMA, Pi~of~Sky, DFN, ATLAS Telescopes, High Time
  Resolution~Universe Survey, RIMAS, RATIR, SKA South~Africa/MeerKAT
  Collaborations, AstroSat Cadmium Zinc Telluride~Imager Team, AGILE Team, 1M2H
  Team, Las Cumbres~Observatory Group, MAXI Team, TZAC Consortium, SALT Group,
  Euro~VLBI Team, and Chandra~Team at~McGill University (Abbott B.~P.\~et al.),
\newblock Multi-messenger Observations of a Binary Neutron Star Merger,
\newblock {\em Astrophys.\ J.} 848 (2017) L12

\bibitem{MultiMess2}
LIGO Scientific, Virgo, Fermi-GBM, and INTEGRAL Collaborations (Abbott B.~P.\
  et~al.),
\newblock Gravitational Waves and Gamma-Rays from a Binary Neutron Star Merger:
  GW170817 and GRB 170817A,
\newblock {\em Astrophys.\ J.} 848 (2017) L13

\bibitem{MultiMess3}
D.~A.\ Coulter, R.~J.\ Foley, C.~D.\ Kilpatrick, M.~A.\ Drout, A.~L.\ Piro,
  B.~J.\ Shappee, M.~R.\ Siebert, J.~D.\ Simon, N.\ Ulloa, D.\ Kasen, B.~F.\
  Madore, A.\ Murguia-Berthier, Y.-C.\ Pan, J.~X.\ Prochaska, E.\ Ramirez-Ruiz,
  A.\ Rest, and C.\ Rojas-Bravo,
\newblock Swope Supernova Survey 2017a (SSS17a), the optical counterpart to a
  gravitational wave source,
\newblock {\em Science} 358 (2017) 1556--1558

\bibitem{MultiMess4}
A.\ Murguia-Berthier, E.\ Ramirez-Ruiz, C.~D.\ Kilpatrick, R.~J.\ Foley, D.\
  Kasen, W.~H.\ Lee, A.~L.\ Piro, D.~A.\ Coulter, M.~R.\ Drout, B.~F.\ Madore,
  B.~J.\ Shappee, Y.-C.\ Pan, J.~X.\ Prochaska, A.\ Rest, C.\ Rojas-Bravo,
  M.~R.\ Siebert, and J.~D.\ Simon,
\newblock A Neutron Star Binary Merger Model for GW170817/GRB 170817A/SSS17a,
\newblock {\em Astrophys.\ J.} 848 (2017) L34

\bibitem{OstrogradskiProblem}
R.~P.\ Woodard,
\newblock Avoiding Dark Energy with $1/R$ Modifications of Gravity,
\newblock {\em Lect.\ Notes Phys.} 720 (2007) 403--433, [astro-ph/0601672]

\bibitem{SphSymmEff}
A.\ Alonso-Bardaj\'{\i}, D.\ Brizuela, and R.\ Vera,
\newblock An effective model for the quantum Schwarzschild black hole,
\newblock {\em Phys.\ Lett.\ B} 829 (2022) 137075, [arXiv:2112.12110]

\bibitem{HigherCov}
M.\ Bojowald and E.~I.\ Duque,
\newblock Emergent modified gravity: Covariance regained,
\newblock {\em Phys.\ Rev.\ D} 108 (2023) 084066, [arXiv:2310.06798]

\bibitem{DiracHamGR}
P.~A.~M.\ Dirac,
\newblock The theory of gravitation in Hamiltonian form,
\newblock {\em Proc.\ Roy.\ Soc.\ A} 246 (1958) 333--343

\bibitem{ConsRinehart}
C.\ Blohmann, M.\ Schiavina, and A.\ Weinstein,
\newblock A Lie-Rinehart algebra in general relativity, [arXiv:2201.02883]

\bibitem{MOND1}
M.\ Milgrom,
\newblock A modification of the Newtonian dynamics-Implications for galaxies,
\newblock {\em Ap.\ J.} 270 (1983) 371--383

\bibitem{MOND2}
S.~S.\ McGaugh and W.\ De~Blok,
\newblock Testing the hypothesis of modified dynamics with low surface
  brightness galaxies and other evidence,
\newblock {\em Ap.\ J.} 499 (1998) 66

\bibitem{TEVES}
J.~D.\ Bekenstein,
\newblock Relativistic gravitation theory for the modified Newtonian dynamics
  paradigm,
\newblock {\em Phys.\ Rev.\ D} 70 (2004) 083509, [astro-ph/0403694]

\bibitem{TEVES2}
J.~W.\ Moffat,
\newblock Scalar–tensor–vector gravity theory,
\newblock {\em JCAP} 2006 (2006) 004, [gr-qc/0506021]

\bibitem{TEVES3}
C.\ Skordis and T.\ Zlosnik,
\newblock A new relativistic theory for Modified Newtonian Dynamics,
\newblock {\em Phys.\ Rev.\ Lett.} 127 (2021) 161302, [arXiv:2007.00082]

\bibitem{SphSymm}
M.\ Bojowald,
\newblock Spherically Symmetric Quantum Geometry: States and Basic Operators,
\newblock {\em Class.\ Quantum Grav.} 21 (2004) 3733--3753, [gr-qc/0407017]

\bibitem{SphSymmMatter}
A.\ Alonso-Bardaj\'{\i} and D.\ Brizuela,
\newblock Holonomy and inverse-triad corrections in spherical models coupled to
  matter,
\newblock {\em Eur.\ Phys.\ J.\ C} 81 (2021) 283, [arXiv:2010.14437]

\bibitem{SphSymmMatter2}
A.\ Alonso-Bardaj\'{\i} and D.\ Brizuela,
\newblock Anomaly-free deformations of spherical general relativity coupled to
  matter,
\newblock {\em Phys.\ Rev.\ D} 104 (2021) 084064, [arXiv:2106.07595]

\bibitem{SphSymmMinCoup}
A.\ Alonso-Bardaj\'{\i} and D.\ Brizuela,
\newblock Spacetime geometry from canonical spherical gravity

\bibitem{HamRen}
T.\ Lang, K.\ Liegener, and T.\ Thiemann,
\newblock Hamiltonian Renormalisation I: Derivation from Osterwalder-Schrader
  Reconstruction,
\newblock {\em Class.\ Quant.\ Grav.} 35 (2018) 245011, [arXiv:1711.05685]

\bibitem{CW}
M.\ Bojowald and S.\ Brahma,
\newblock Canonical derivation of effective potentials
\newblock (2014), [arXiv:1411.3636]

\bibitem{ColemanWeinberg}
S.\ Coleman and E.\ Weinberg,
\newblock Radiative corrections as the origin of spontaneous symmetry breaking,
\newblock {\em Phys.\ Rev.\ D} 7 (1973) 1888--1910

\bibitem{SphSymmMoments}
G.\ Sims, M.\ D\'{\i}az, K.\ Berglund, and M.\ Bojowald,
\newblock Quasiclassical solutions for static quantum black holes,
  [arXiv:2012.07649]

\bibitem{LoopSchwarz}
R.\ Gambini and J.\ Pullin,
\newblock Loop quantization of the Schwarzschild black hole,
\newblock {\em Phys.\ Rev.\ Lett.} 110 (2013) 211301, [arXiv:1302.5265]

\bibitem{MONDTest}
S.~S.\ McGaugh,
\newblock Novel test of modified Newtonian dynamics with gas rich galaxies,
\newblock {\em Phys.\ Rev.\ Lett.} 106 (2011) 121303

\bibitem{MONDRotation}
M.\ Milgrom and E.\ Braun,
\newblock The rotation curve of DDO 154-A particularly acute test of the
  modified dynamics,
\newblock {\em Ap.\ J.} 334 (1988) 130--133

\end{thebibliography}

\end{document}